\documentclass[pra,twocolumn,,showpacs,amsmath,amssymb,floatfix]{revtex4}

\pdfoutput=1

\usepackage{amsmath}
\usepackage{amssymb}
\usepackage[pdftex]{graphicx}
\usepackage{graphicx}
\usepackage{dcolumn}
\usepackage{bm}
\usepackage{color}

\newcommand{\be}{\begin{equation}}
\newcommand{\ee}{\end{equation}}
\newcommand{\bea}{\begin{eqnarray}}
\newcommand{\eea}{\end{eqnarray}}

\begin{document}

\title{Flat bands, Dirac cones and atom dynamics in an optical lattice}

\author{V. Apaja, M. Hyrk\"as and M. Manninen}

\affiliation{Nanoscience Center, Department of Physics, FIN-40014 University
  of  Jyv\"askyl\"a, Finland}

\date{\today}
 
\begin{abstract}
We study atoms trapped with a harmonic confinement 
in an optical lattice characterized by a flat band and Dirac cones. 
We show that such an optical lattice can be constructed which can be accurately
described with the tight binding or Hubbard models. In the case of fermions 
the release of the harmonic confinement removes fast atoms occupying the 
Dirac cones while those occupying the flat band remain immobile. 
Using exact diagonalization and dynamics we demonstrate that a 
similar strong occupation of the flat band does not happen in bosonic case and
furthermore that the mean field model is not capable for describing the
dynamics of the boson cloud.

\end{abstract}


\maketitle

\section{Introduction}          

Recent development of manipulating cold atoms in optical lattices has
opened new avenues for studying correlation and band structure effects
in a controlled
way\cite{bloch2005,leggett2006,pethick2008,bloch2008,lopezaguayo2010}.
One, two and three-dimensional lattices with a variety of lattice
structures have been proposed, including the kagome
lattice\cite{santos2004,damski2005,wu2007,ruostekoski2009}, which was
originally suggested by Syo\^zi\cite{syozi1951} and studied
intensively in the condensed matter physics mainly due to its magnetic
properties\cite{sachdev1992,greedan2001,kimura2002,balents2010}.  In a
simple tight binding (TB) model the kagome lattice is characterised by
a completely flat band, meaning that the energy is independent of the
wave vector $\bf k$ and that electrons occupying this band are
localized. Flat band Hubbard model is a paradigm for ferromagnetism,
recently reviewed by Tasaki~\cite{tasaki2008}.
 
Kagome lattice is one example of a whole class of two (2D) and three-dimensional (3D)
lattice structures which, in the TB model, produce a flat band\cite{deng2003,miyahara2005}.
In 2D these structures include square and hexagonal lattices and can have one or
several flat bands as well as crossing points where two bands open as circular cones
leading to Dirac fermions, as in graphene\cite{castroneto2009}. 
The Dirac fermions are also known to lead to interesting magnetic properties\cite{meng2010}.
Allowing also the p-states in each lattice site to be occupied even more freedom to
create flat bands and Dirac cones is obtained. Wu {\it et al}\cite{wu2007} 
have studied the honeycomb
(graphene) optical lattice with p-electrons and demonstrated the atom localization 
due the flat band of the system.

\begin{figure}[h!]
\includegraphics[width=0.30\columnwidth]{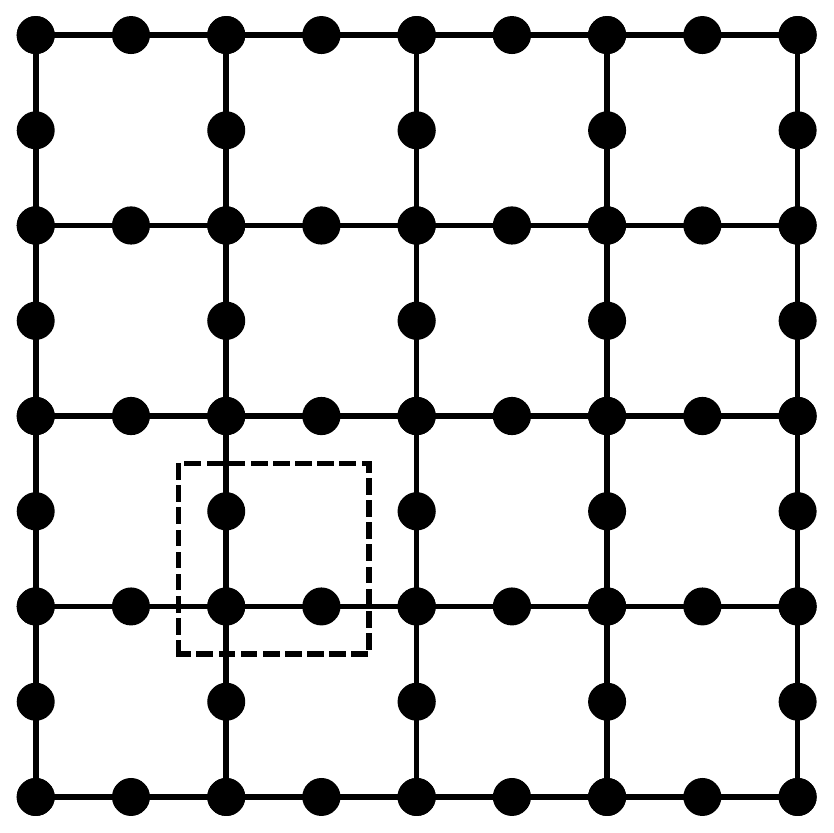} 
\includegraphics[width=0.45\columnwidth]{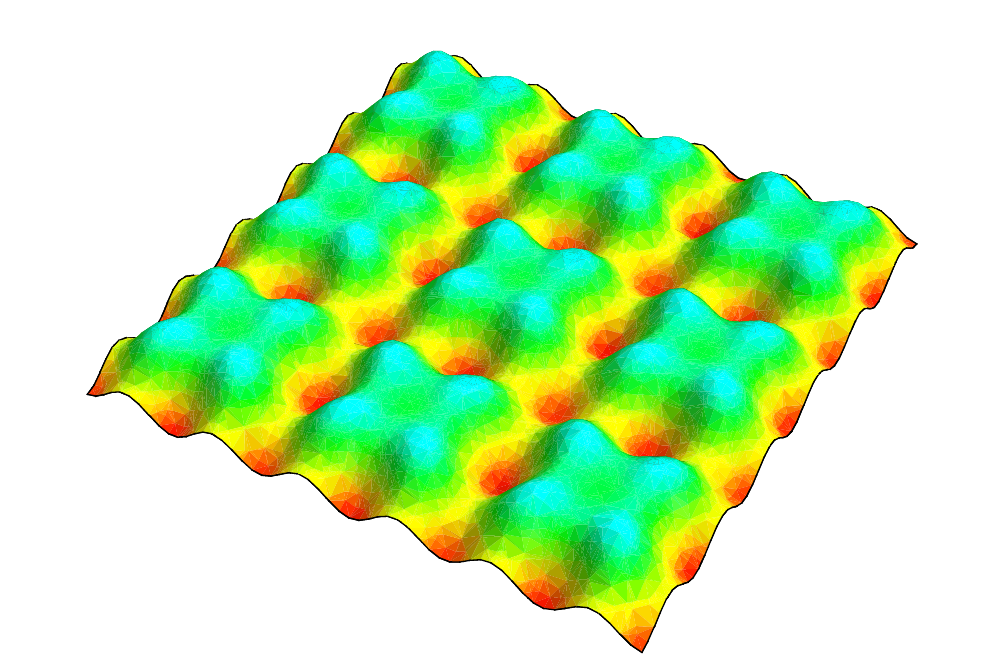} 
\includegraphics[width=0.30\columnwidth]{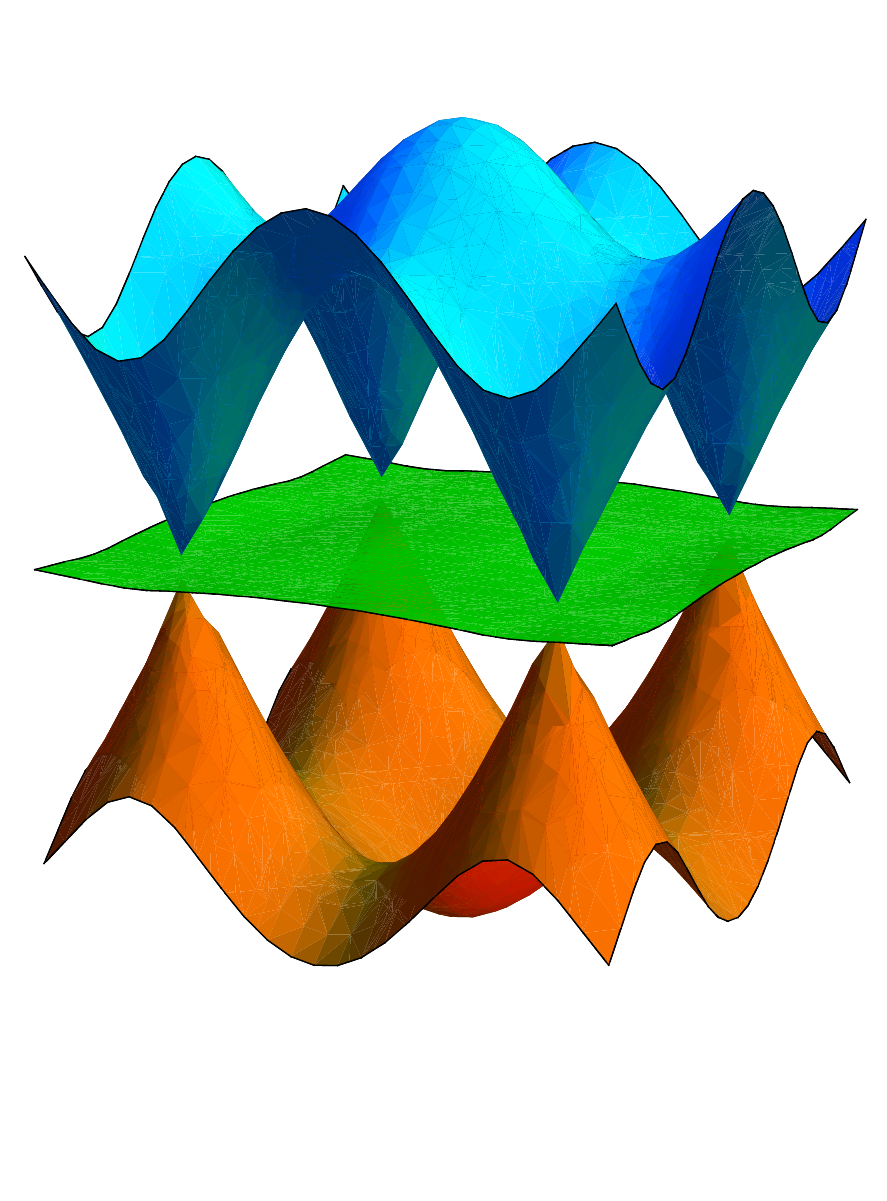} \hskip0.1\columnwidth 
\includegraphics[width=0.35\columnwidth]{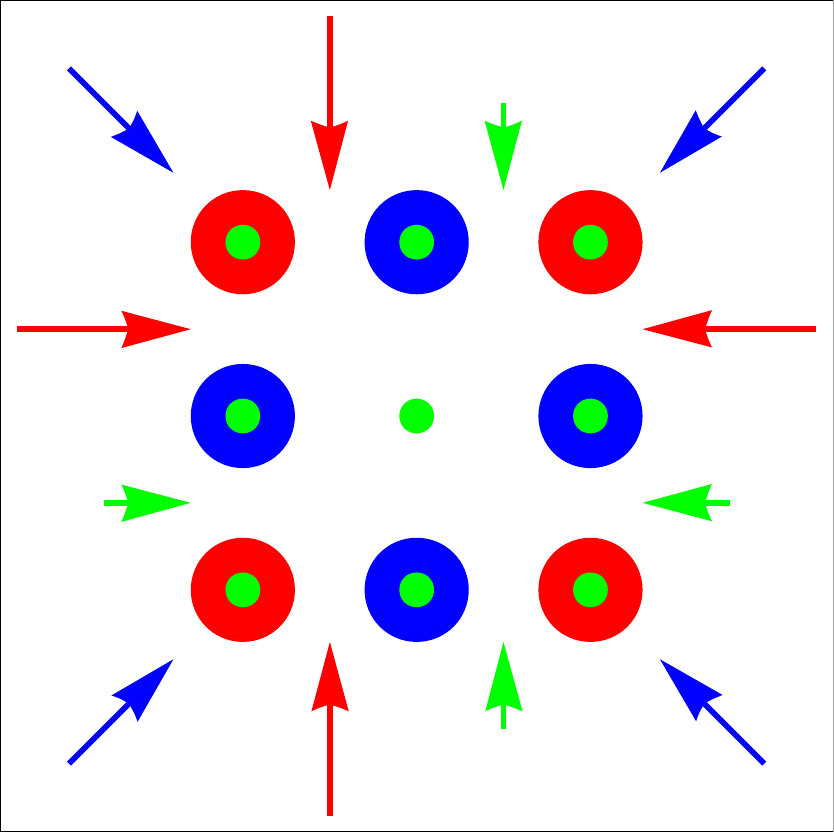} 
\caption{{\it Up left}: Edge-centered square lattice and its unit cell
shown with a dashed line.
{\it Up right}: Pptimized potential, Eq. (3), with equally deep minima shown with red color.
{\it Down left}: Three lowest bands corresponding to the potential of Eq. (3). The cusps are
at the corners of the Brillouin zone.
{\it Down right}: Arrows show the suggested directions of the laser fields
causing potentials like $\cos^2 x$, $\cos^2(x+y)$ and $\cos^2(2x)$, 
corresponding to minima at the sites of the red, blue and green dots,
respectively.
}
\label{lat}
\end{figure}

One of the simplest flat band lattices is the edge-centered square
lattice with three atoms in a unit cell\cite{aoki1996}, depicted in
Fig.~\ref{lat}.  In the TB model this lattice has three bands, the
center band is flat and the upper and lower bands meet the flat band
at the corners of the Brilloun zone where Dirac cones open. This means
that at the same energy the particle can have an infinite effective
mass (flat band) or zero effective mass (Dirac fermions). Recently
Shen {\it et al.}  showed that these massless Dirac fermions can
exhibit perfect all-angle Klein tunneling \cite{shen2010}.  This is
the lattice of interest in the present work.

We will first show that laser field can be used to create a lattice
which accurately produces a flat band and Dirac cones meeting at the
same energy.  The band structure is then fold to a TB model and for
interacting atoms to a Hubbard model. In the case of fermions with
only one spin state (spinless fermions) the Hubbard model equals to
the noninteracting fermions since the Pauli exclusion principle
prevents two similar fermions to occupy the same lattice point and
on--site interaction has no effect. This provides us a simple way to
study a dynamical problems, for example, what happens when a harmonic
confinement, keeping the atoms in the central region of the lattice,
is removed.

In the case of bosons the many-particle problem of the Hubbard model
becomes more complicated. In this case we solve the problem
with the Hubbard model for a small system of only three particles. However,
this already demonstrates that (i) the boson system behaves differently than
the fermion system and (ii) that frequently used mean field models
(e.g. Gross-Pitaevskii), where the boson system is described with only one
single particle wave function, fails to describe the dynamics correctly.

The results show that, when the harmonic confinement is removed, part of
the atoms fly away fast while part remain stuck in the immobile flat band
states. We anticipate that this kind of experiment can be made in the
near future using both bosonic and fermionic atoms.

The edge-centered square lattice with
three atoms in a unit cell is illustrated in Fig. \ref{lat}. 
We consider fist a simple TB model with only one state per lattice site, assume only
nearest neighbour hopping and neglect the so-called differential overlap
between neighbours. The band structure can be easily solved\cite{ashcroft1976}
by diagonalizing a $3\times 3$ matrix for each 2D wave vector ${\bf q}=(q_x,q_y)$,
resulting to energy levels
\be
\begin{array}{l}
\epsilon_1({\bf q})=-t\sqrt{4+2\cos(q_x a)+2\cos(q_y a)}\\
\epsilon_2({\bf q})=0\\
\epsilon_3({\bf q})=t\sqrt{4+2\cos(q_x a)+2\cos(q_y a)},\\
\end{array}
\ee
where $t$ is the strength of the hopping integral between nearest
neighbours and $a$ the lattice constant. The corresponding wave vectors $v_i=(v_{\rm c},v_{\rm e1},v_{\rm e2})$,
where the three components refer to the corner and edge sites in the unit cell,
are ($s=\sqrt{4+e^{iq_xa}+e^{-iq_xa}+e^{iq_ya}+e^{-iq_ya}}$)
\be
\begin{array}{l}
v_{1}({\bf q})=(-s,\quad 1+e^{-iq_ya},\quad 1+e^{-iq_xa})\\
v_{2}({\bf q})=(0,\quad 1+e^{iq_xa},\quad 1+e^{iq_ya})\\
v_{3}({\bf q})=(s,\quad 1+e^{-iq_ya},\quad 1+e^{-iq_xa}).\\
\end{array}
\ee
The energy bands consist of a flat band at zero energy and two bands symmetrically
below and above the flat band. All bands meet at points at the corners of the 
Brillouin zone, $q_x=q_y=\pi/a$. The band structure is shown in Fig. \ref{lat}.
The wave function of the flat band has zero amplitude at the corner point of
the square lattice. This is the reason of the flatness: The particles in these states do not
hop to the neighboring sites and can be viewed as localized. The effective band mass
of the particles is thus infinite.

Close to the zero energy the two other bands form "Dirac cones", and the 
velocity of the particles is independent of the energy, 
$\epsilon_{1,3}(k_x-\pi/a,k_y-\pi/a)\approx \mp t a\sqrt{k_x^2+k_y^2}/4=\mp t a k/4$,
where $k$ is the distance from the point of the cusp. 
The linear dispersion relation means a zero effective mass. Interestingly,
in this lattice we have at the same energy region particles with zero effective
mass and infinite effective mass. 
Note that the cones appear to be symmetric to rather high energies indicating that
most of the particles in the lowest and uppermost bands have nearly equal velocities.

Next we will consider how such a simple TB lattice could be constructed 
for cold atoms in an optical lattice. We do not go in the details of the
interactions between atoms and laser fields\cite{bloch2005,leggett2006,pethick2008},
but simply assume that the effective potential caused by the standing electromagnetic
waves can be described with trigonometric functions ($\cos^2( x)$ or equivalently $\cos(2 x)$).
Clearly a square lattice can be made as $-\cos^2(a x/\pi)-\cos^2(a y/\pi)$,
having minima at lattice sites $R=(na,ma)$. Minima at the edge centers can
be constructed with functions $\cos^2(a(x+y)/2)+\cos^2(a(x-y)/2$.
However, with a linear combination of these two it is not possible to construct
proper potential barriers between corner end edge sites and an additional 
shorter wave length function of $-\cos^2(2ax/\pi)-\cos^2(2ay/\pi)$ has to be added.
The potential optimized to give the band structure similar to that of the TB
band structure is 
\be
\begin{array}{rl}
V_{\rm OL}(x,y)=&V_0[-0.9(\cos^2[a x/\pi]+\cos^2[a y/\pi])\\
&+0.496(\cos^2[a(x+y)/2]+\cos^2[a(x-y)/2])\\
&+(\cos^2[2ax/\pi]+\cos^2[2ay/\pi])],\\
\end{array}
\ee
where $V_0$ gives the energy scale (see below).
The resulting potential is shown in Fig. \ref{lat}. 
The lowest three bands of the band structure 
(computed using a real space mesh of $26\times 26$ points)
are shown in Fig. \ref{lat} and they
are nearly identical to those of the simple TB model.
The parameter $V_0$ is related to the mass of the atoms $m$ and
the lattice constant $a$ as $V_0\approx 497\hbar^2/(2ma^2)$ (where the
numerical factor 497 comes from fitting of the band structure to that
of the TB model).

The overall width of the three lowest bands is $7.65\hbar^2/(2ma^2)$ 
while the width of the flat band
is only $0.11\hbar^2/(2ma^2)$, meaning that the dispersion of the flat band 
is only 1.5 \% of the total width of the three bands.
The potential of Eq. (3) has, of course, an infinite number of other bands,
but there is a huge gap of   $140\hbar^2/(2ma^2)$ 
between the three lowest bands and the next band.
With this example we have demonstrated that a proper combination of cosine
potentials can produce a band structure which, with a high accuracy, 
gives similar energy bands as the simple TB model.
Figure \ref{lat} shows schematically the laser arrangement 
needed for making such an optical lattice.

We assume that the atoms trapped by the optical lattice contact with a
repulsive short range potential and assume the interaction is so strong that
only one atom is allowed in each lattice site. The many-particle problem
then reduces to the Hubbard Hamiltonian. We first consider 
fermionic atoms with only one spin state (spinless fermions).
In this case the many-particle problem is noninteracting, since 
the Pauli exclusion principle already prevents two atoms to occupy 
the same site. 

In the actual experiments the optical lattice is superimposed on a
harmonic confinement which localizes the particles (with repulsive interaction)
in the central region of the lattice. 
Putting enough atoms in the lattice the harmonic confinement 
guarantees that in the central region the occupation of each lattice site
is nearly one. This means that all the three bands (which are bend in the
harmonic confinement) are filled at the central region of the harmonic confinement.

Our interest is to see what happens when the harmonic confinement is removed 
but the optical lattice is kept in place.  This dynamical problem is easily
solved for $N$ spinless fermions. The procedure is as follows: 
We solve the TB model in a large
finite lattice with a harmonic confinement and fill the lowest $N$ single particle states.
This is the fermionic ground state $\Psi_0(0)$ at time zero. 
Next we remove the harmonic confinement
and expand the states with the confinement in terms of the states
$\Psi^0$ calculated without the confinement. The time dependence 
follows then from the time dependences of each individual state as
\be
\Psi_0(t)=\sum_j C_j e^{-iE_j t/\hbar}\Psi_j^0(0),
\ee
where $E_j$ is the energy of the many-particle state $j$.
In the case of spinless fermions where each state is a Slater determinant
of single particle states the time dependence further reduces to the
time dependences of the single particle states.

\begin{figure}[h!]
\begin{center}
$\begin{array}{cc}
\includegraphics[width = 0.45\columnwidth]{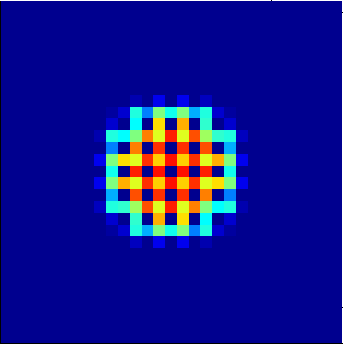} &
\includegraphics[width = 0.45\columnwidth]{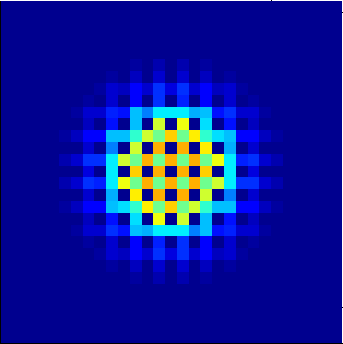} \\
\mbox{\textbf{(a)}  t = 0} & \mbox{\textbf{(b)} t = 2.5} \\ [0.5cm]
\includegraphics[width = 0.45\columnwidth]{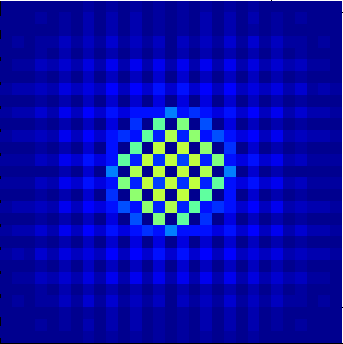} &
\includegraphics[width = 0.45\columnwidth]{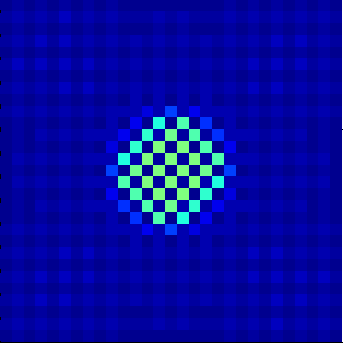} \\
\mbox{\textbf{(c)} t = 5} & \mbox{\textbf{(d)} t = 7.5} \\ [0.5cm]
\multicolumn{2}{c}{\includegraphics[width =0.45\columnwidth]{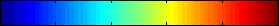}} \\
\mbox{\quad 0} & \mbox{1 \;\;}
\end{array}$
\end{center}
\caption{Time evolution of 50 fermionic atoms when the armonic confinement
is removed. (a) shows the initial state with the harmoic confinement of strength
$U_{\rm h}=0.4$ (in units of $\hbar^2/2ma^2$), and (d) the 'final' state when atoms
trapped by the flat band stay localized while other otoms have flown away. 
}
\label{tdfer}
\end{figure}

In practice we solved up to 100 atoms in a lattice of $25\times 25$
unit cells. We use TB units with the hopping parameter $t=1$ and the nearest 
neighbour distance $b=a/2=1$. The harmonic confinement adds to diagonal
terms to the TB Hamiltonian matrix: Placing the center of the finite lattice at the origin,
the harmonic confinement changes the energy level of a lattice site by
$U_{\rm h} r^2$, where $U_{\rm h}$ is the strength of the harmonic confinement and 
$r$ the distance of the lattice site from the origin (in units of the lattice constant).
Figure \ref{tdfer} shows an example of such a simulation.
The Harmonic confinement localizes the atom cloud at the center of the 
finite lattice. The maximum density is at the edge sites of the unit cell with occupation
about 0.8, while at the corner sites it is about 0.6. When the harmonic 
confinement is removed, the cloud fast expands, Fig. \ref{tdfer} (b) and (c),
but part of the atoms stay trapped at the edge sites, (d), forming a lattice
of localized atoms. The occupation of the edge states remains close to 0.5.

The explanation of the observed dynamics is a direct result of the single-particle
properties of noninteracting fermions: Those occupying the bands 1 and 3 
are mobile and fly fast away (at the dirac cones even with the same velocity),
while atoms occupying the flat band are immobile and stay in place.  
Notice that the corner states of the unit cell are emptied since the flat band 
does not have any amplitude in these sites.

The bosonic many-particle problem is usually simpler than the fermionic case
due to the symmetry of the wave function. In our case, however, this is not the case.
Even if we assume an infinitely strong contact interaction the bosonic case
remains a true many-particle problem, although the
boson and fermion problems are related\cite{ouvry2009}. 
A common approach is to use a mean
field model for the Hubbard Hamiltonian (this is closely related to the
Gross-Pitaevskii\cite{gross1961,pitaevskii1961} model): The interaction
term $U(\hat n_i-1)\hat n_i$ is replaced with $U(\langle n_i\rangle-1) \hat n_i$ 
(where $\hat n_i$ is the occupation number operator). 
The first approximation is to
assume that all the bosons occupy the same quantum state determined 
by the external confinement. 

Taking the interaction $U$ very large, we prevent the occupation of any lattice
site to be much larger than one. In the Harmonic confinement we can then 
get initial atom density to be qualitatively similar to that of the fermionic case
and one would expect that expanding the self-consistent bosonic ground state
in terms of the single particle band states also the flat band is markedly occupied. 
However, this is not the case. The occupation 
of the flat band state remains less than 2 \% even when the interaction strength
$U$ is extrapolated to infinity. This means that when the confinement 
was removed, practically all the atoms of the cloud escape fast.

\begin{figure}[h!]
\includegraphics[width=0.3\columnwidth]{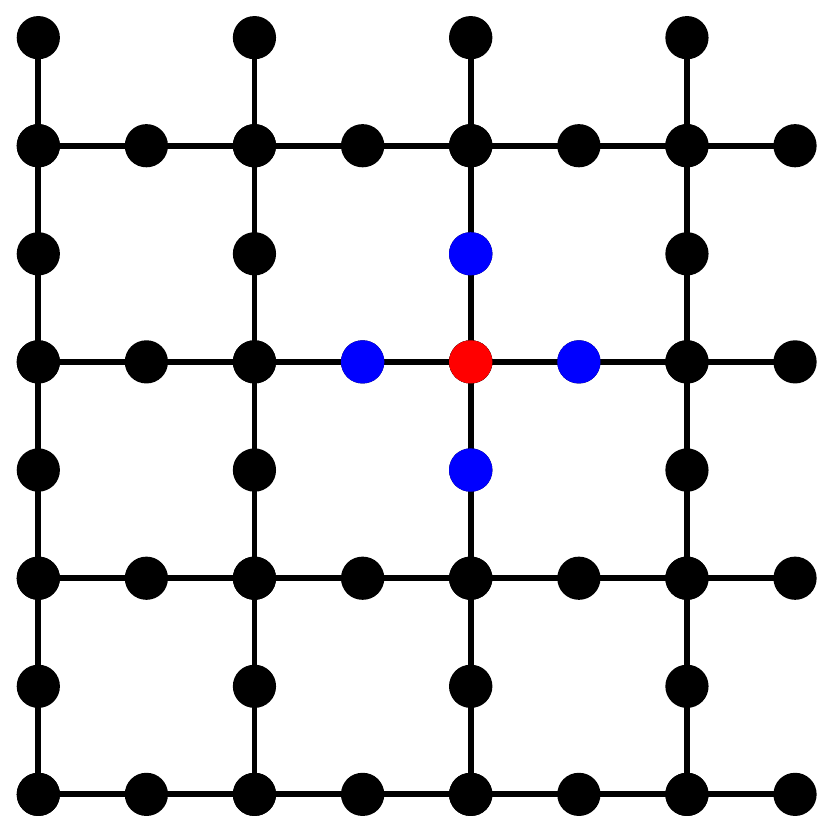} 
\includegraphics[width=0.5\columnwidth]{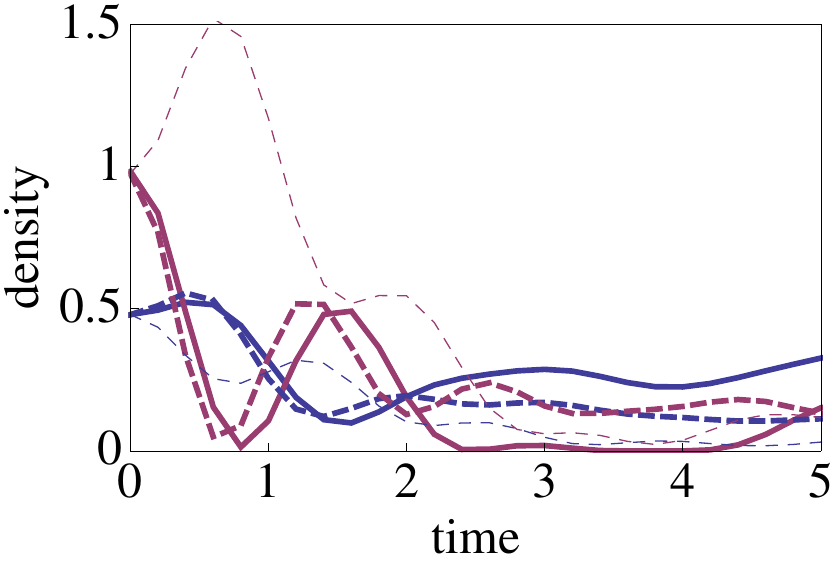} 
\caption{{\it Left}: Finite lattice used, wiht periodic boundary conditions, 
for solving the Hubbard model for three bosons or fermions. The initial
potential at the cites denoted by red and blue were $-20t$ and $-5t$,
respectively.
{\it Right}: Time dependece of the occupation of red and blue sites after the 
potential of these sites were put to zero. Solid lines are results for fermions and
the thick dashed lines for bosons. The thin dashed lines are results for 
a single particle with the same initial density as the exact boson result,
indicating that the mean field solution fails in this case. The time is in units of 
$\hbar/t$
}
\label{hubbard}
\end{figure}

In order to understand if the boson system behaves truly differently than the 
fermion system, or if it is the mean field approach which is not appropriate
for our system, we performed a small scale exact diagonalization of the 
Hubbard Hamiltonian. The test size was $3\times 3$ unit cells with periodic
boundary condition, i.e. 27 sites altogether as shown in Fig. \ref{hubbard}.
We assumed infinitely strong repulsive interaction between the atoms
($U=\infty$) which reduced the boson basis to states having at most
one atom in a lattice site. Intially the atoms were localized in a small region
with strong attractive potentials in five adjacent sites as shown in Fig. \ref{hubbard}.
By diagonalizing the Hubbard Hamiltonian for three bosons (and fermions)
with and without the attractive potentials at five sites, the time evolution
could be solved exactly using Eq. (4). 

The time evolution is shown in Fig. \ref{hubbard} for bosons and fermions.
In the case of fermions the results is qualitatively similar to that of larger systems
described above. After a short time the atom density at the corner site (red)
decreases to nearly zero while at the edge sites (blue) it stays quite large, 
averaging to a value of about 0.3 after initial oscillations. 
In the case of bosons the initial time dependence is very similar,
but soon both center and edge sites average to the same value of about 0.1,
which is close to the average filling of the lattice $3/27=1/9$.
Figure \ref{hubbard} also shows the time evolution of a single state which initially
had the same density distribution as the bosonic case. Clearly, this time evolution
is qualitatively different indicating that a mean field (single state) approximation
for the bosonic system can not describe correctly the time evolution in the 
present case.

In conclusion, we have shown that using cosine functions a rather simple lattice
can be constructed, which has a flat band and Dirac cones meeting at the same energy.
The three lowest band of the lattice can be accurately described with a simple
tight binding model with only one state per site. 
Confing atoms with a harmonic confinement in such a lattice fermion atoms stay
trapped in the flat band states even when the harmonic confinement is
removed.

The case of bosons is more complicated and requires further study.
Using the Hubbard model with infinitely strong interaction ($U\rightarrow \infty$)
we demonstrated that the trapping of the bosons is not as strong
as in the case of fermions. Similar result was obtained with the mean field
approximation for bosons, although comparison to the exact solution
of the Hubbard model indicated that the mean field model fails to describe the
boson dynamic correctly in the system studied.

\bibliography{apaja}

\end{document}